\documentclass[12pt]{article} 
\usepackage{amsmath}
\usepackage{amssymb}
\usepackage{bm}
\usepackage{siunitx}
\usepackage{cite}
\usepackage{lmodern}

\usepackage[a4paper]{geometry} 
\begin{document}	

\title{Variation of Vorticity Gradient and Formation of Sunspots}
\author{Chen Haibin, Wu Rong\thanks{wurong2@mail3.sysu.edu.cn}}
\date{}
\maketitle
	   	   
\section*{\centering Abstract}  

Based on the rotating turbulent thermal convection model and using the rotating equivalent temperature assumption and new convection criterion, this paper analyzed the repression of the vorticity gradient on the heat transport and explained that the formation of sunspots originated from the variation of the vorticity gradient in the solar troposphere.
  
\section*{Introduction}

The most prominent feature on the solar surface is sunspot, where the dark area is caused by low temperature. The temperature of solar photosphere is close to $6000 \mathrm{K}$, and  sunspot is $4100 \sim 4200 \mathrm{K}$. The temperature of umbra is $4000 \sim 4500 \mathrm{K}$, and penumbra is about $5500 \mathrm{K}$. There are scattered, non-flaky granules called umbral dots in the umbra, which are about $30 \%$ smaller in size than normal granules in the photosphere.

Compared with photospheric model, it can be found that the density and pressure of the gas in umbra are lower than that of the solar quiet region at the same geometrical height. The opacity of the umbra is lower than that of the surrounding photosphere, and the geometric depth of the sunspot is deeper than that of the photosphere, so the sunspot corresponds to groove and depression on the solar surface. The Wilson effect indicates that when a circular sunspot rotates with the sun to the edge of the sun, the penumbra near the sun's center is much narrower than the penumbra near the sun's edge, and sunspot can be regarded as a dish sag \cite{1}.

Sunspots are mostly represented by pairs or groups and accompanied by strong magnetic fields. The larger the sunspot, the larger the magnetic field and the longer the lifetime of sunspot. Therefore, it is considered that the sunspot is essentially a locally strong magnetic field on the solar surface, and the low temperature and dark color are the results of magnetic field restraining convection \cite{2}.

We have found that vorticity transport affects the convection and its criterion in rotating turbulent thermal convection in studies of solar differential rotation and granulation \cite{3,4}. In the solar quiet regions, fluids larger than the critical size are in forced convection state and those smaller than the critical size are in natural convection state. The critical size can be used as the feature length of thermal convection. The vorticity and vorticity gradient, which affect the critical size and then the efficiency of heat transport in thermal convection, are powerful candidates for the formation mechanism of sunspots.

\section{Criterion of Rotating Turbulent Thermal Convection}

\subsection{Rotating equivalent pressure and rotating equivalent temperature}

When the characteristic time of thermal convection is much less than the rotational period of the planet, it can be assumed that the expansion of the thermal convection fluid cell is statistically isotropic. When isotropic expansion occurs, additional pressure is generated by the rotation of the fluid cell, and the rotational kinetic energy changes due to the work done by the rotating equivalent pressure. The rotational equivalent pressure is related to the rotational speed of the fluid cell and can be expressed as
\begin{equation}\label{Eq.1}
	p_\Omega = k_\Omega \rho \Omega^2 a^2   ,
\end{equation}
where $\rho$ is density of the fluid cell, $\Omega$ is the rotational speed of the fluid cell, $a$ is the amount related to the size of the fluid cell, $k_\Omega$ is a constant related to the shape of the fluid cell. For example, for a cylindrical fluid cell with a radius of $a$, a height of $2 a$, and a rotational speed of $\Omega$, $k_\Omega = \frac{1}{6}$. Since the average pressure is independent of the rotational direction, the rotational speed is taken as a scalar. In fluid mechanics, vorticity is used instead of rotational speed to describe the rotation of the fluid cell for ease of calculation. According to $\omega = 2 \Omega$, $k_\omega = k_\Omega$, the rotating equivalent pressure of the fluid cell is expressed by vorticity as
\begin{equation}\label{Eq.2}
	p_\omega = \frac{1}{4} k_\omega \rho \omega^2 a^2   .
\end{equation}

It is easier to illustrate the rotating equivalent pressure by analogy with ideal gas law $p = \frac{R_\mathrm{m}}{M_\mathrm{m}} \rho T$, in which $R_\mathrm{m}$ is the molar gas constant, $M_\mathrm{m}$ is the gas molar mass and $T$ is gas temperature. Suppose that
\begin{equation}\label{Eq.3}
	p_\omega = \frac{R_\mathrm{m}}{M_\mathrm{m}} \rho T_\omega   ,
\end{equation}
the rotating equivalent pressure of the fluid cell also corresponds to a rotating equivalent temperature, which is
\begin{equation}\label{Eq.4}
	T_\omega = \frac{1}{4} \frac{M_\mathrm{m}}{R_\mathrm{m}} k_\omega \omega^2 a^2   .
\end{equation}
The concept of rotating equivalent temperature can be verified on a cylindrical fluid cell. When the fluid cell does not transport the rotational kinetic energy generated by viscosity with the outside, the work done by the rotating equivalent pressure of the adjacent fluid on the fluid cell should be equal to the increase of the rotational kinetic energy of the fluid cell. This process is similar to the adiabatic compression of an ideal gas.

The rotational kinetic energy of a cylindrical fluid cell with mass of $M$, radius of $a$, height of $2 a$, and vorticity of $\omega$ is
\begin{equation}\label{Eq.5}
	E_\omega = \frac{1}{16} M \omega^2 a^2   ,
\end{equation}
and it can also be expressed by the rotating equivalent temperature as
\begin{equation}\label{Eq.6}
	E_\omega = \frac{3}{2} \frac{M}{M_\mathrm{m}} R_\mathrm{m} T_\omega   ,
\end{equation}
Its form is similar to the internal energy of monatomic gas with degree of freedom $i = 3$, and the molar heat capacity at constant volume is $\frac{3}{2} R_\mathrm{m}$. 

If the viscosity is neglected, the angular momentum of the fluid cell is conserved during isotropic expansion, and the relationship between $\omega$ and $a$ is
\begin{equation}\label{Eq.7}
	\frac{\mathrm{d} \omega}{\omega} = - 2 \frac{\mathrm{d} a}{a}   .
\end{equation}
The mass of fluid cell is conserved, then the relationship between $a$ and $\rho$ is
\begin{equation}\label{Eq.8}
	\frac{\mathrm{d} a}{a} = - \frac{1}{3} \frac{\mathrm{d} \rho}{\rho}  .
\end{equation}
By differentiating $T_\omega$, we get 
\begin{equation}\label{Eq.9}
	\mathrm{d} T_\omega = T_\omega \left( 2 \frac{\mathrm{d} \omega}{\omega} + 2 \frac{\mathrm{d} a}{a} \right) = \frac{2}{3} T_\omega \frac{\mathrm{d} \rho}{\rho}  ,
\end{equation}
the relationship between temperature and density is the same as that of ideal gas with $i=3$ during adiabatic expansion. 

The work done by the outside on the rotating equivalent pressure of a cylindrical fluid cell during isotropic expansion
\begin{equation}\label{Eq.10}
	- p_\omega \mathrm{d} V = - \frac{R_\mathrm{m}}{M_\mathrm{m}} \rho T_\omega \left( - V \frac{\mathrm{d} \rho}{\rho} \right) = \frac{3}{2} \frac{M}{M_\mathrm{m}} R_\mathrm{m} \mathrm{d} T_\omega = \mathrm{d} E_\omega     .
\end{equation}
It is proved that the work done by the rotating equivalent pressure of the adjacent fluid on the fluid cell is equal to the increase of the rotational kinetic energy of the fluid cell and this model is similar to the adiabatic compression of an ideal gas. The assumption of rotational equivalent temperature is reasonable and provides a new method for calculating rotating turbulent thermal convection. 

\subsection{Convection Criterion Affected by Rotating Equivalent Temperature}

The existence time of solar granules is much shorter than the solar rotational period. The disequilibrium of pressure caused by the inconsistency of vorticity between the fluid cell and the adjacent fluid can be neglected. The expansion process of fluid cell is close to isotropy statistically. The thermal properties of rotating fluid cell are fully displayed and the convection criterion needs to consider the effect of rotating equivalent temperature.

The Schwarzschild criterion for irrotational fluid is
\begin{equation}\label{Eq.11}
	\left| \frac{\mathrm{d} T}{\mathrm{d} l} \right| _{\mathrm{rd}} > \left| \frac{\mathrm{d} T}{\mathrm{d} l} \right| _{\mathrm{ad}} ,
\end{equation}
where $\left( \frac{\mathrm{d} T}{\mathrm{d} l} \right) _{\mathrm{rd}}$ is the real temperature gradient of the fluid, $\left( \frac{\mathrm{d} T}{\mathrm{d} l} \right) _{\mathrm{ad}}$ is the adiabatic temperature gradient and can be expressed as
\begin{equation}\label{Eq.12}
	\left( \frac{\mathrm{d} T}{\mathrm{d} l} \right) _{\mathrm{ad}} = \left( 1-\frac{1}{\gamma} \right) T \frac{\mathrm{d} p}{p \mathrm{d} l} ,
\end{equation}
or
\begin{equation}\label{Eq.13}
	\left( \frac{\mathrm{d} T}{\mathrm{d} l} \right) _{\mathrm{ad}} = \left( \gamma - 1 \right) T \frac{\mathrm{d} \rho}{\rho \mathrm{d} l} .
\end{equation}
Considering the effect of rotating equivalent temperature, the criterion of rotating turbulent thermal convection is
\begin{equation}\label{Eq.14}
	\left| \frac{\mathrm{d} T + \mathrm{d} T_\omega}{\mathrm{d} l} \right| _{\mathrm{rd}} > \left| \frac{\mathrm{d} T + \mathrm{d} T_\omega}{\mathrm{d} l} \right| _{\mathrm{ad}} ,
\end{equation}
in which the rotating equivalent adiabatic temperature gradient is
\begin{equation}\label{Eq.15}
	\left( \frac{\mathrm{d} T_\omega}{\mathrm{d} l} \right) _{\mathrm{ad}} = \frac{2}{3} T_\omega \frac{\mathrm{d} \rho}{\rho \mathrm{d} l} = T_\omega \left( 2 \frac{\mathrm{d} \omega }{\omega \mathrm{d} l} + 2 \frac{\mathrm{d} a }{a \mathrm{d} l} \right) .
\end{equation}
For a single fluid cell, because of the mass conservation, $\frac{\mathrm{d} a }{a \mathrm{d} l} = - \frac{1}{3} \frac{\mathrm{d} \rho }{\rho \mathrm{d} l}$, then we get
\begin{equation}\label{Eq.16}
	\frac{\mathrm{d} \omega}{\omega \mathrm{d} l} = \frac{2}{3} \frac{\mathrm{d} \rho }{\rho \mathrm{d} l} ,
\end{equation}
which is the same as the relationship between vorticity and density of inviscid rotating fluid cell in the process of isotropic expansion. So the rotating equivalent adiabatic temperature gradient can also be described by vorticity gradient, that is, 
\begin{equation}\label{Eq.17}
	\left( \frac{\mathrm{d} \omega}{\omega \mathrm{d} l} \right) _\mathrm{ad} = \frac{2}{3} \frac{\mathrm{d} \rho }{\rho \mathrm{d} l} .
\end{equation}

Since the two temperature gradients are different but interact with each other, the convection criterion does not need to strictly satisfy $\left| \frac{\mathrm{d} T}{\mathrm{d} l} \right| _{\mathrm{rd}} > \left| \frac{\mathrm{d} T}{\mathrm{d} l} \right| _{\mathrm{ad}}$ and $\left| \frac{\mathrm{d} T_\omega}{\mathrm{d} l} \right| _{\mathrm{rd}} > \left| \frac{\mathrm{d} T_\omega}{\mathrm{d} l} \right| _{\mathrm{ad}}$, that is, convection does not necessarily need to be driven simultaneously by temperature gradient and rotating equivalent temperature gradient.  When one of them is satisfied, the convection criterion $\left| \frac{\mathrm{d} T + \mathrm{d} T_\omega}{\mathrm{d} l} \right| _{\mathrm{rd}} > \left| \frac{\mathrm{d} T + \mathrm{d} T_\omega}{\mathrm{d} l} \right| _{\mathrm{rd}}$ can be satistfied by limiting the size of the fluid cell. According to the critical condition of convection criterion $\left| \frac{\mathrm{d} T + \mathrm{d} T_\omega}{\mathrm{d} l} \right| _{\mathrm{rd}} = \left| \frac{\mathrm{d} T + \mathrm{d} T_\omega}{\mathrm{d} l} \right| _{\mathrm{ad}}$, the critical size of fluid cell can be calculated as
\begin{equation}\label{Eq.18}
	a^2_\mathrm{ad} = - \frac
	{T \left[ \frac{\mathrm{d} T}{T \mathrm{d} l} - \left( \gamma - 1 \right) \frac{\mathrm{d} \rho}{\rho \mathrm{d} l} \right]}
	{\frac{k_\omega M_m}{2 R_m} \omega^2 \left( \frac{\mathrm{d} \omega}{\omega \mathrm{d} l} - \frac{2}{3} \frac{\mathrm{d} \rho}{\rho \mathrm{d} l} \right)} .
\end{equation}
The discriminant of convection criterion can be discussed in several cases:

Case 1: When $\left| \frac{\mathrm{d} T}{T \mathrm{d} l} \right| < \left| \left( \gamma - 1 \right) \frac{\mathrm{d} \rho}{\rho \mathrm{d} l} \right|$ and $\left| \frac{\mathrm{d} \omega}{\omega \mathrm{d} l} \right| < \left| \frac{2}{3} \frac{\mathrm{d} \rho}{\rho \mathrm{d} l} \right|$, natural convection can not occur;

Case 2: When $\left| \frac{\mathrm{d} T}{T \mathrm{d} l} \right| > \left| \left( \gamma - 1 \right) \frac{\mathrm{d} \rho}{\rho \mathrm{d} l} \right|$ and $\left| \frac{\mathrm{d} \omega}{\omega \mathrm{d} l} \right| < \left| \frac{2}{3} \frac{\mathrm{d} \rho}{\rho \mathrm{d} l} \right|$, the convection is driven by the temperature gradient, and the description of convection criterion by the fluid cell size is $a < a_\mathrm{ad}$; 

Case 3: When $\left| \frac{\mathrm{d} T}{T \mathrm{d} l} \right| < \left| \left( \gamma - 1 \right) \frac{\mathrm{d} \rho}{\rho \mathrm{d} l} \right|$ and $\left| \frac{\mathrm{d} \omega}{\omega \mathrm{d} l} \right| > \left| \frac{2}{3} \frac{\mathrm{d} \rho}{\rho \mathrm{d} l} \right|$, the convection is driven by the vorticity gradient, and the description of convection criterion by the fluid cell size is $a > a_\mathrm{ad}$; 

Case 4: When $\left| \frac{\mathrm{d} T}{T \mathrm{d} l} \right| > \left| \left( \gamma - 1 \right) \frac{\mathrm{d} \rho}{\rho \mathrm{d} l} \right|$ and $\left| \frac{\mathrm{d} \omega}{\omega \mathrm{d} l} \right| > \left| \frac{2}{3} \frac{\mathrm{d} \rho}{\rho \mathrm{d} l} \right|$, the convection is driven by both temperature gradient and vorticity gradient, and the fluid cell size of the convection criterion is unlimited. 

The deduction of the above critical size of fluid cell and discriminant of convevtion criterion is valid when $\frac{\mathrm{d} T}{T \mathrm{d} l}$ and $\frac{\mathrm{d} \omega}{\omega \mathrm{d} l}$ have the same sign as $\frac{\mathrm{d} \rho}{\rho \mathrm{d} l}$. When there is a different sign between them, special cases need to be analysed specifically. 

\section{Formation Mechanism of Sunspots}
\subsection{Repression of vorticity gradient on convection in solar troposphere}

The solar thermal convection is driven by the temperature gradient which meets $\left| \frac{\mathrm{d} T}{T \mathrm{d} l} \right| > \left| \left( \gamma - 1 \right) \frac{\mathrm{d} \rho}{\rho \mathrm{d} l} \right|$ radially, and the vorticity distribution generally satisfies $\left| \frac{\mathrm{d} \omega}{\omega \mathrm{d} l} \right| < \left| \frac{2}{3} \frac{\mathrm{d} \rho}{\rho \mathrm{d} l} \right|$. In Case 2 of convection criterion, heat transport provides energy for convection, and vorticity transport absorbs energy from convection. The larger the size of the fluid cell, the stronger the absorption capacity of convection energy by vorticity transport. When $a = a_\mathrm{ad}$, the energy provided by heat transport is equal to that absorbed by vorticity transport. Critical size $a_\mathrm{ad}$ can also be used as a feature size of convection fluid cells.

From the calculation of the critical size $a_\mathrm{ad}$ of the fluid cells, the vorticity $\omega$ and the relative vorticity gradient $\frac{\mathrm{d} \omega}{\omega \mathrm{d} l}$ will influence the critical size $a_\mathrm{ad}$, and $a_\mathrm{ad}$ decreases as $\omega^2 \left| \frac{\mathrm{d} \omega}{\omega \mathrm{d} l} - \frac{2}{3} \frac{\mathrm{d} \rho}{\rho \mathrm{d} l} \right|$ increases. $a_\mathrm{ad}$ represents the limitation of the vorticity gradient on the thermal convection Reynolds number $Re$, which increases with increase of $a_\mathrm{ad}$. Reynolds number could influence the heat transport capacity of fluid, thereby change the temperature gradient. Intuitively, when the energy transport speed is basically invariant, the larger $a_\mathrm{ad}$ is, the smaller the convection limitation is, which is equivalent to increasing the heat energy transport generated by larger fluid cells, that is, $\left| \frac{\mathrm{d} T}{T \mathrm{d} l} - \left( \gamma - 1 \right) \frac{\mathrm{d} \rho}{\rho \mathrm{d} l} \right|$ increases, and both of $\frac{\mathrm{d} T}{T \mathrm{d} l}$ and $T$ decrease. 

A simple heat transfer model of the solar troposphere can be established by combining the heat conductivity of ideal gas with the mixing length theory. The heat conductivity of ideal gas is
\begin{equation}\label{Eq.19}
	k = \frac{1}{3} \rho \bar{v} \bar{\lambda} \frac{C_V}{M_m} ,
\end{equation}
in the mixing length theory, $\bar{\lambda}$ can be considered as the mixing length and $\bar{v}$ as the turbulent pulsation velocity. Dimensional analysis shows that under other invariable conditions, $\bar{v} \bar{\lambda}$ increases with the increase of Reynolds number $Re$. Therefore, in the solar troposphere, the heat conductivity $k$ of turbulence increases with the increase of the critical size $a_\mathrm{ad}$ of the fluid cells. Heat flux $\phi$ in thermal convection can be expressed as
\begin{equation}\label{Eq.20}
	\phi = - k T \left( \frac{\mathrm{d} T}{T \mathrm{d} l} - \left( \gamma - 1 \right) \frac{\mathrm{d} \rho}{\rho \mathrm{d} l} \right) ,
\end{equation}
which shows that with the invariant heat flux, the larger $a_\mathrm{ad}$, the larger $k$ and thus the smaller $\left| \frac{\mathrm{d} T}{T \mathrm{d} l} - \left( \gamma - 1 \right) \frac{\mathrm{d} \rho}{\rho \mathrm{d} l} \right|$. As the heat flux decreases with decreasing temperature, the temperature gradient changes more significantly.

To sum up, in Case 2 of the convection criterion, when $\omega$ and $\frac{\mathrm{d} \omega}{\omega \mathrm{d} l}$ increase, $a_\mathrm{ad}$ decreases, then $\frac{\mathrm{d} T}{T \mathrm{d} l}$ and $T$ decrease. Similarly, when $\omega$ and $\frac{\mathrm{d} \omega}{\omega \mathrm{d} l}$ decrease, $a_\mathrm{ad}$ increases, then $\frac{\mathrm{d} T}{T \mathrm{d} l}$ and $T$ increase.

\subsection{Variation of vorticity gradient and formation of sunspots}

Most of the solar convection zone is in Case 2 of convection criterion most of the time. Affected by solar magnetic field or other factors, west winds and east winds form on the solar surface. When a west wind occurs, $\omega$ and $\frac{\mathrm{d} \omega}{\omega \mathrm{d} l}$ increase as the linear velocity $v_\theta$ of the fluid on the solar surface increases,  and then $a_\mathrm{ad}$ increases, resulting in decreases of $\frac{\mathrm{d} T}{T \mathrm{d} l}$ and $T$ , therefore the temperature of the west wind is low. Similarly, the temperature of the east wind is high.

The temperature changes in the solar convection zone will lead to further effects. In the west wind region, when the temperature at the top of the solar troposphere significantly decreases, the gravity of the fluid is greater than the buoyancy, resulting in downwelling. The local pressure decreases, the surrounding gas is absorbed, and the rotational speed of the surrounding gas increases under the action of Coriolis force, forming a large-scale cyclone. The angle between the direction of rotation of the cyclone and the direction of rotation of the sun is less than 90 degrees. The formation of the cyclone is similar to typhoon generation, but wind direction in the core area is top-down. Cyclone generation can significantly increase local $\omega$ and $\frac{\mathrm{d} \omega}{\omega \mathrm{d} l}$, which will significantly reduce $a_\mathrm{ad}$ and convection intensity. Energy and momentum transport of the convection are further reduced and the temperature is decreased considerably, forming positive feedback. The cyclones can be maintained spontaneously to form observable stable low-temperature zones, which are consistent with the observed phenomena of smaller granules size, lower temperature and dish sag of sunspots.

The formation of the solar low-temperature cyclone changes the velocity of the surrounding gas, blowing strong east winds and strong west winds on the solar surface. Strong west winds can induce the generation of new sunspots, forming sunspot groups. Strong east winds can alter the convection criteria and produce an energy explosion.

In the area of strong east winds caused by sunspots, $\frac{\mathrm{d} \omega}{\omega \mathrm{d} l}$ is significantly reduced. When $\frac{\mathrm{d} \omega}{\omega \mathrm{d} l} < \frac{2}{3} \frac{\mathrm{d} \rho}{\rho \mathrm{d} l}$, i.e. $\left| \frac{\mathrm{d} \omega}{\omega \mathrm{d} l} \right|< \left| \frac{2}{3} \frac{\mathrm{d} \rho}{\rho \mathrm{d} l} \right|$, the convection criterion changes from Case 2 to Case 4. In Case 4, the fluid cell size is unlimited, and the convection is driven by both temperature gradient and vorticity gradient. The larger the fluid cell size, the stronger the driving power received. With the feature size of the fluid cells is in the order of magnitude similar to the thickness of the region with $\left| \frac{\mathrm{d} \omega}{\omega \mathrm{d} l} \right|< \left| \frac{2}{3} \frac{\mathrm{d} \rho}{\rho \mathrm{d} l} \right|$, the relative temperature gradient limit is close to $\left( \gamma - 1 \right) \frac{\mathrm{d} \rho}{\rho \mathrm{d} l}$, and the temperature of the solar surface rises sharply. It can be found that, when $\frac{\mathrm{d} \omega}{\omega \mathrm{d} l}$ crosses the critical value of $\frac{2}{3} \frac{\mathrm{d} \rho}{\rho \mathrm{d} l}$, the severe changes of the convection structure and relative temperature gradient cause dramatic changes in density, which can lead to explosions or eruptions. This feature is consistent with observed flares and prominences.

The area where a sunspot is located blocks the upward transport of energy from the lower layer, causing temperature to rise at its bottom and around it. The increase of vorticity and relative vorticity gradient directly caused by the low-temperature cyclone corresponding to the sunspot will cause the abnormal velocity distribution at the cyclone boundary, which will lead to the abnormal convection state at the cyclone boundary. This may be the reason why sunspots can not exist permanently and related phenomena need further study. 

Sunspots are accompanied by strong magnetic fields, which can be explained by planetary dynamos \cite{5,6}. Conductive fluid convection inside the planet generates planetary magnetic fields, and plasma cyclones of sunspots generate magnetic fields of sunspots. Because the cyclones are fast and last for a long time, the magnetic fields generated by sunspots are very strong.

\section{Conclusion}

In this paper, the additional pressure generated by the rotation of the fluid cell in the rotating turbulent thermal convection is explained by the assumption of the rotating equivalent temperature, and a new convection criterion related to vorticity and vorticity gradient is obtained. Combined with the actual situation of the solar troposphere, the generation of sunspots, flares, prominences and sunspot magnetic fields are explained. The main conclusions are as follows:

(1) Rotation of fluid cells under isotropic expansion creates the additional rotating equivalent pressure and rotating equivalent temperature, which can influence the convection criterion.The solar troposphere is mainly in Case 2 of the convection criterion, in which the convection is driven by the temperature gradient, and the fluid cells less than the critical size are in natural convection. When the vorticity and vorticity gradient increase, the critical size of the fluid cells decreases, and then the temperature and temperature gradient decrease. Similarly, when the vorticity and vorticity gradient decrease, the critical size of the fluid ells increases, and then the temperature and temperature gradient increase. 

(2)When the west wind appears on the solar surface, the vorticity and vorticity gradient of the fluid on the solar surface increase, resulting in the decrease of temperature and temperature gradient, so the temperature of the west wind is low. Similarly, when the east wind appears on the sun surface, the temperature of the west wind is high. When the large area at the top of the solar west wind  lower the temperature significantly, gravity is greater than buoyancy, resulting in a large top-down cyclone. The vorticity and vorticity gradient of the fluid increase greatly, leading to a sharp drop in temperature and temperature gradient, forming a positive feedback. The cyclones can be maintained spontaneously to form observable stable low-temperature zones, which are consistent with the observed phenomena of smaller granules size, lower temperature and dish sag of sunspots.

(3)The formation of the solar low-temperature cyclone changes the velocity of the surrounding gas, blowing strong east winds and strong west winds on the solar surface. Strong west winds can induce the generation of new sunspots, forming sunspot groups. Strong east winds can change the convection criterion from Case 2 to Case 4. The severe convection and sudden change of density lead to explosions or eruptions, which is consistent with the observed flare and prominences.

\bibliographystyle{IEEEtran}
\bibliography{reference.bib}

 \end{document}